\documentclass[a4paper]{article}
 \advance \textheight by \topskip
\usepackage{graphicx}
\newcommand{\R}{{\rm I\! R}}
\def\Z{{\mathchoice {\hbox{$\sf\textstyle Z\kern-0.4em Z$}}
{\hbox{$\sf\textstyle Z\kern-0.4em Z$}}
{\hbox{$\sf\scriptstyle Z\kern-0.3em Z$}}
{\hbox{$\sf\scriptscriptstyle Z\kern-0.2em Z$}}}}
\def\C{{\mathchoice {\setbox0=\hbox{$\displaystyle\rm C$}\hbox{\hbox
to0pt{\kern0.4\wd0\vrule height0.9\ht0\hss}\box0}}
{\setbox0=\hbox{$\textstyle\rm C$}\hbox{\hbox
to0pt{\kern0.4\wd0\vrule height0.9\ht0\hss}\box0}}
{\setbox0=\hbox{$\scriptstyle\rm C$}\hbox{\hbox
to0pt{\kern0.4\wd0\vrule height0.9\ht0\hss}\box0}}
{\setbox0=\hbox{$\scriptscriptstyle\rm C$}\hbox{\hbox
to0pt{\kern0.4\wd0\vrule height0.9\ht0\hss}\box0}}}}
\begin {document}
\title {
The First Birkhoff Coefficient and the Stability of 2-Periodic Orbits on Billiards}
\author{ Sylvie OLIFFSON KAMPHORST and S\^onia PINTO DE CARVALHO
}
\date{Departamento de Matem\'atica ICEx UFMG\\
Caixa Postal 702, 30123-970 Brazil\\
syok@mat.ufmg.br, sonia@mat.ufmg.br}
\maketitle

\begin{abstract}
In this work we address the question of proving the stability of elliptic 2-periodic orbits for strictly convex billiards. Eventhough it is part of a widely accepted belief that ellipticity implies stability, classical theorems show that the certainty of stability relies upon more fine conditions. We present a review of the main results and general theorems and describe the procedure to fullfill the supplementary conditions for strictly convex billiards.

\end{abstract}

\section{Introduction}

Let $\alpha$ be a plane, closed, regular and strictly convex curve. 
The billiard problem on $\alpha$ consists in the free motion of a 
point particle in the plane region enclosed by $\alpha$,
with unitary velocity and being reflected elastically at the impacts with the boundary.
The trajectories are polygonals in the region.

\begin{center}
\includegraphics[width=6cm]{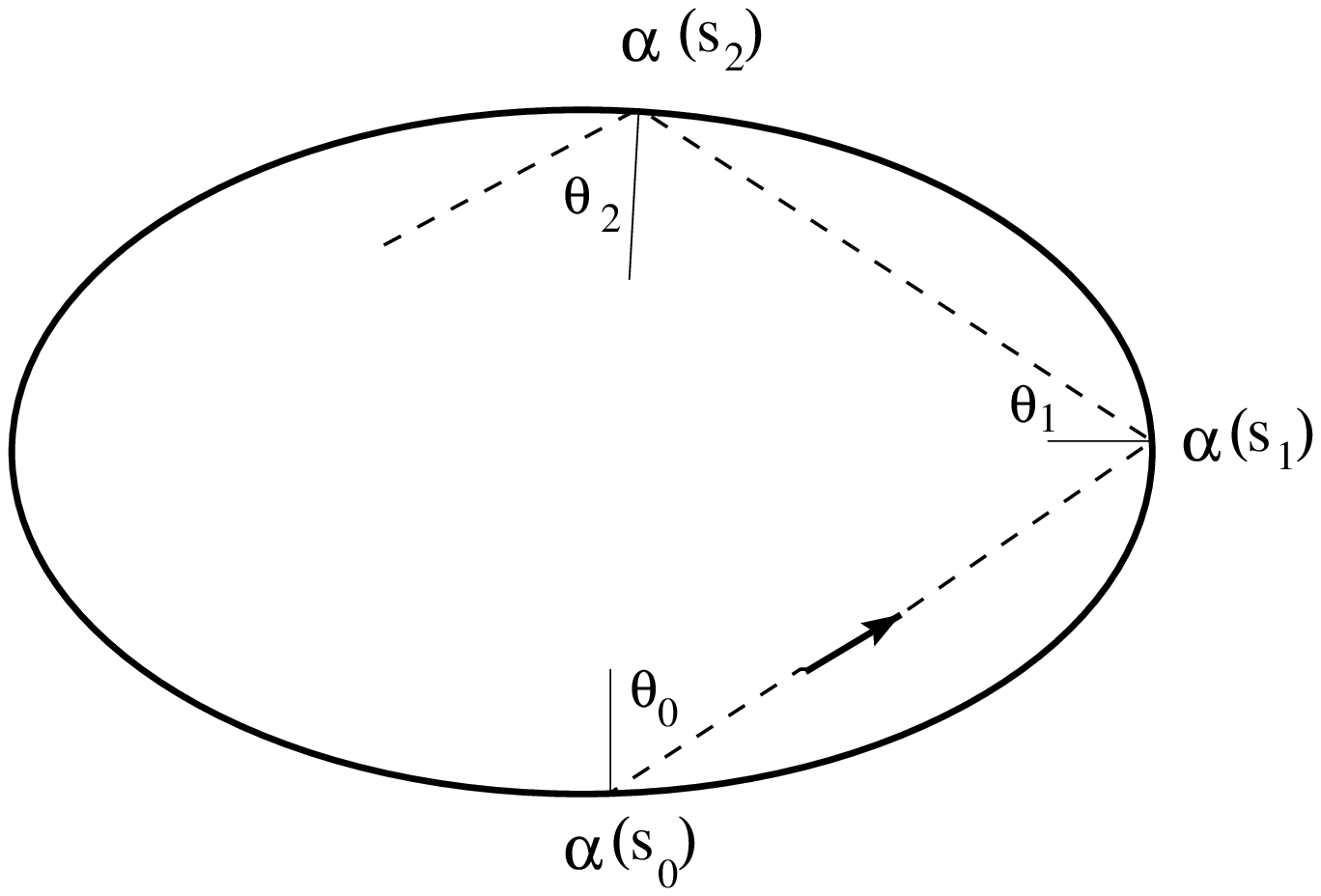}
\end{center} 

The motion is completely determined  by the point of reflection at
$\alpha$ and the direction of motion immediately after each reflection.
For instance, the arclength parameter $s$, which locates the point of reflection, and the tangential component of the momentum $p=\sin\theta$, where $\theta$ is the angle 
between the direction of motion and the normal to the boundary at the reflection point, 
describe the system. Good introductions to billiards can be found in
\cite{bir}, \cite{mar}, \cite{has}, \cite{kat}, \cite{ams},  \cite{str} or \cite{tab}.

The billiard model defines a map $T$ which to each $(s,p)$ in the annulus 
${\cal A} = [0,L)\times (-1,1)$, representing the pair impact coordinate and direction of 
motion, associates the next impact and direction:
$$
\begin{array}{cccc}
 T :& {\cal A} & \to &{\cal A} \cr
    & (s,p) & \longmapsto & (S(s,p),P(s,p))
\end{array}
$$

Since the particle can travel along the same polygonal 
in both senses, the problem is time-reversing and the inverse map $T^{-1}$ is well defined.

The derivative of $T$ at $(s,p)$ is implicitly calculated and is given by the formulae:
\begin{eqnarray}
\frac{\partial S}{\partial s}&=&\frac{l(s,p)-R(s)\cos\theta(p)}
{R(s)\cos\theta(P)}  \nonumber\\
\frac{\partial S}{\partial p}&=&\frac{l(s,p)}
{\cos\theta(p)\cos\theta(P)} \label{eq:deriv} \\
\frac{\partial P}{\partial s}&=&\frac{l(s,p)-R(s)\cos\theta(p)-R(S)\cos\theta(P)}
{R(s)R(S)} \nonumber\\
\frac{\partial P}{\partial p}&=&\frac{l(s,p)-R(s)\cos\theta(P)}
{R(S)\cos\theta(p)} \nonumber
\end{eqnarray}
 where $S$ stands for $S(s,p)$ and $P$ for $P(s,p)$, $l(s,p)$ is the distance between the two consecutive impacts points $\alpha(s)$ and $\alpha(S)$, $R$ is the radius of curvature of $\alpha$ and $\cos \theta (p) = \sqrt{1-p^2}$ is the normal component of the momentum.
 
If $\alpha$ is a $C^k$ curve, $k\geq 2$, the billiard model gives rise to a discrete two-dimensional $C^{k-1}$ area preserving dynamical system, whose orbits are given by
$${\cal O}(s, p)=\{ T^j(s, p), j \in \Z \}\subset {\cal A}. $$ 

A billiard has no fixed points. However, given $n\geq 2$, Birkhoff's Theorem states that $T$ has at least two different orbits of period $n$ which will be fixed points of $T^n$. The linearization of $T^n$ at any of these fixed points, say $(s,p)$, gives the linear 
area preserving map $DT^n_{(s, p)}$, which has a fixed point at the origin $(0,0)$. 
According to the eigenvalues of this linear map, the fixed point $(s,p)$ is classified as: 
hyperbolic if the eigenvalues of  are $\mu$ and $\frac{1}{\mu}$, $\mu\in\R$, $\mu\neq\pm 1$,
elliptic if the eigenvalues are $\mu=e^{i\gamma}$ and $\overline\mu$,
$\mu^2\neq 1$ or parabolic if the eigenvalues are 1 or -1. 

In the hyperbolic case, the Hartman-Grobman Theorem (see, for instance, \cite{kat} or 
\cite{jac}) assures that, on a neighbourhood of the fixed point $(s,p)$, the dynamical 
behaviour of $T^n$ is the same as the dynamical behaviour of $DT^n_{(s, p)}$ on a 
neighbourhood of the origin. So, $(s,p)$ is an unstable fixed point of $T^n$ and 
$\{(s,p),T(s,p),...,T^{n-1}(s,p)\}$ is an unstable periodic orbit of $T$.
In this case, the instability of the equilibrium of the linear map $DT^n_{(s, p)}$ 
implies the local instability of the periodic orbit for the complete map $T$. 

In the elliptic case, the linear map $DT^n_{(s, p)}$ is 
a rotation: the origin is surrounded by closed invariant circles and is a stable 
equilibrium. However, this beautiful behaviour may not be inherited by the map $T^n$, 
as it can be seen in the examples on the figure bellow.
For both of them, the fixed point is linearly elliptic. On the left side, the non linear map 
exhibit invariant closed curves surrounding the fixed point, which is then stable. For the 
non linear map on the right, no invariant curves can be observed and the fixed point seems 
to be unstable.
\footnote{Even more surprising is the example given by Anosov and 
Katok in \cite{ano} of an ergodic area-preserving map of the disc $|z|<1$, with an elliptic fixed 
point at $z=0$. The ergodicity implies that the fixed point is unstable. This example 
does not represent a billiard map and we don't know if there are any billiards with this 
property.}

\begin{center}
\includegraphics[bb = 50 200 450 400,width=5cm]{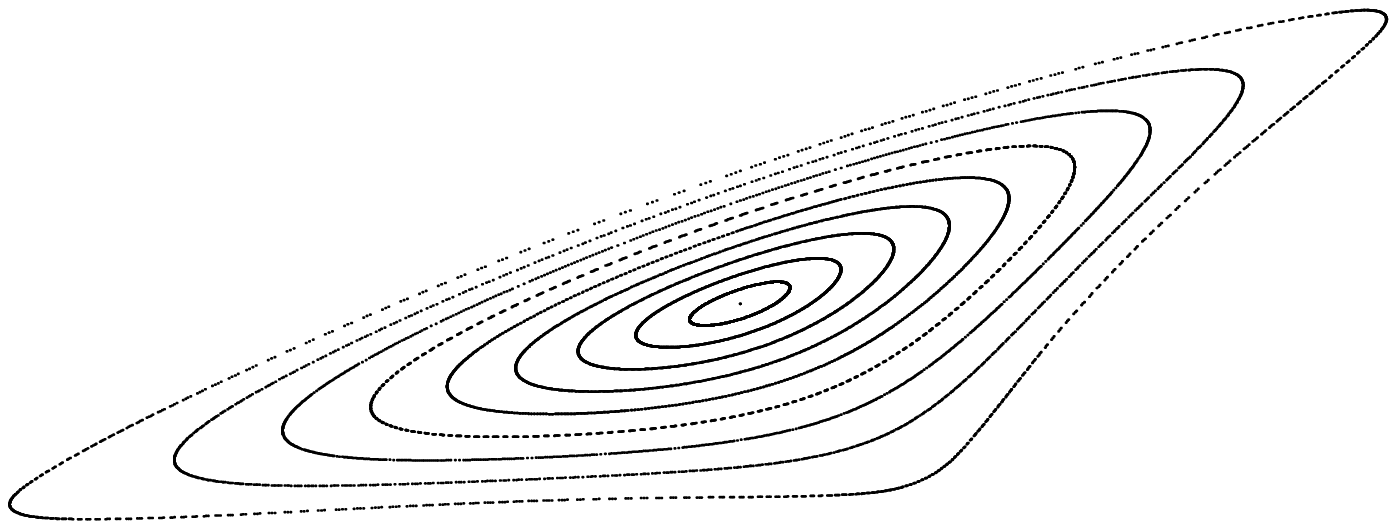}
\hskip 3truecm
\includegraphics[bb = 0 20 400 220,width=5cm]{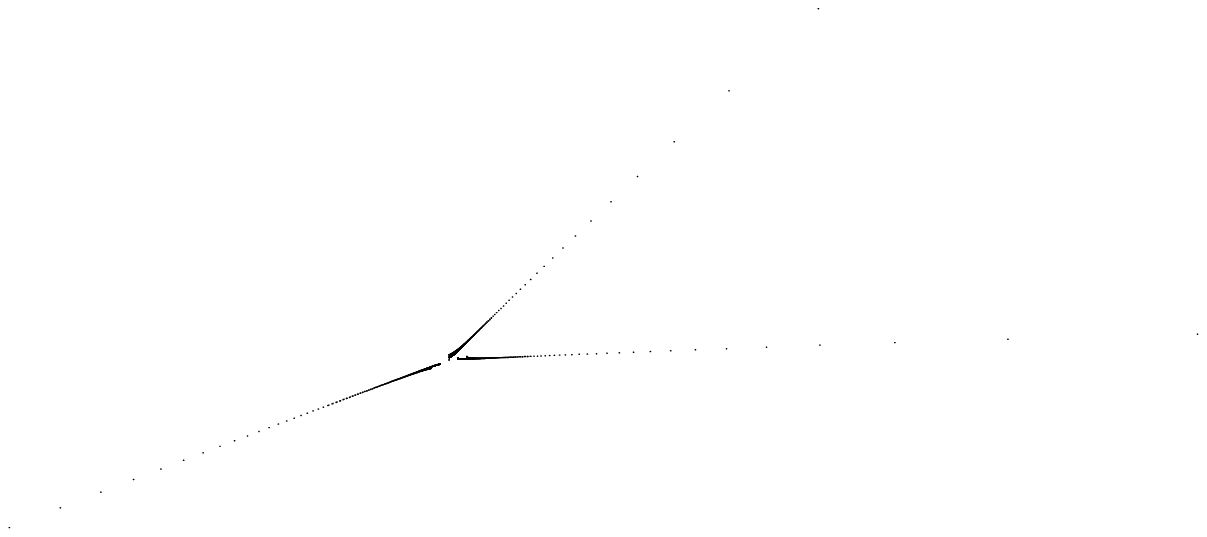}
\end{center}

Moreover, it is not even clear if the pictures above, obtained by numerical simulations, 
correspond to the true behaviour of the maps. In fact, Dias Carneiro and ourselves 
\cite{dia} have proved that any {\bf $C^1$ strictly convex} billiard map with an elliptic 2-periodic orbit can be approached by billiard maps with a 2-periodic orbit surrounded by 
closed invariant curves, ie, with a stable orbit. We guess this result can be extended
to any period. Therefore, because of natural numerical round off errors, one can not be 
sure that the simulation corresponds to the actual billiard at and not to a very close one. 

As a consequence, in the elliptic case, a more carefull approach is needed and higher order terms must be taken into account to assure the local stability of periodic orbits.
A classical way to handle this problem is to use the Birkhoff Normal Form and Moser's Twist  Theorem \cite{sie}.

In what follows we explain how this can be performed and applied to the billiard map in 
the case of 2-periodic elliptic orbits. We have employed the software \textregistered Maple to calculate the necessary data and all the worksheets are available at http://www.mat.ufmg.br/$\sim$syok/papers. We then apply the results to two special classes of billiards.

Related works are Hayli, Dumont, Moulin-Ollagnier, Strelcyn \cite{hay} and Moeckel \cite{moe}. The first authors studied the stability of periodic orbits for a family of Robnik's billiards. The last author studied the generic behaviour of the first Birkhoff coefficient for one-parameter families of conservative maps.

\section{Non linear analysis and the local stability of elliptic orbits}

Let $T$ be an area preserving map with a $n$-periodic orbit
$\{(0,0),T(0,0),...,T^{n-1}(0,0)\}$.
We will assume that the map is $C^k$ with $k\geq 4$. In the case of the billiard map, 
this is equivalent to assume that the curve $\alpha$ is at least $C^5$.

The map $T^n$ can then be expanded in Taylor Form up to order 3 in a neighbourhood of its 
fixed point $(0,0)$, 
\begin{equation}
T^n(s,p)=(a_{10}s+a_{01}p+a_{20}s^2+...+a_{03}p^3,
b_{10}s+b_{01}p+b_{20}s^2+...+b_{03}p^3)+{\cal O}(|(s,p)|^4).
\label{eq:taylor}
\end{equation}
If the fixed point is elliptic, with eigenvalues $\mu=\cos\gamma+i\sin\gamma$ and 
$\bar{\mu}$, by means of a complex linear area preserving coordinate change which 
diagonalizes the linear part, the map $T^n$ can be written as 
\begin{equation}
z \mapsto \mu (z + c_{20} z^2+c_{11}z\overline z+
c_{02}\overline z^2+c_{30} z^3+c_{21} z^2\overline z
+c_{12}z\overline z^2+c_{03}\overline z^3) +{\cal O}(|z|^4). 
\label{eq:comp}
\end{equation}  

If $\mu^j\neq 1$, $j=1,2,3$ or $4$ we say that $\mu$ is non resonant and 
an analytic coordinate change brings the map into its convergent Birkhoff Normal Form
$$
 z \mapsto  {\rm e}^{i(\gamma + \tau_1 |z|^2)} z  +{\cal O}(|z|^4)
=  \mu z+i\mu \tau_1 z|z|^2+{\cal O}(|z|^4) .
$$
The first Birkoff coefficient $\tau_1$ is given by
\begin{equation}
\tau_1=\Im (c_{21}) +\frac{\sin \gamma}{\cos\gamma - 1} 
\left( 3 \left|  \! \, c_{20}\, \!  \right| ^{2} 
+ \frac {2\,\cos\gamma - 1 }{2\,\cos\gamma  + 1}
\left|  \! \,{c_{02}}\, \!  \right| ^{2}
\right )
\label{eq:tau1}
\end{equation}
where $\Im (c_{21})$ stands for the imaginary part of $c_{21}$.

The calculations leading to expression (\ref{eq:tau1}) are standard (\cite{hay}, 
\cite{moe}) and can also be easily performed using symbolic programming.
\footnote{http://www.mat.ufmg.br/$\sim$syok/papers/3NormalForm.}

By Moser's Twist Theorem if the first Birkhoff coefficient $\tau_1$
is not zero there are $T^n$-invariant curves surrounding the fixed point and therefore it 
is stable. We have that each point of the n-periodic orbit is contained in an
open set, called an island, homeomorphic to a disk and invariant under $T^n$.
Each island contains $T^n$-invariant curves surrounding the periodic point.
So, the n-periodic orbit of $T$ is stable.

\section{Elliptic 2-periodic orbits of convex billiards}
\label{sec:3}

Any closed regular strictly convex $C^2$ curve $\alpha$ has at least two diameters, 
characterized by points with parallel tangents and equal normal lines (like the axis of an 
ellipse).  The motion along each one of them corresponds to a 2-periodic trajectory 
for the billiard map associated to $\alpha$.

It is easy to prove that the longest of these diameters, if isolated, 
corresponds to a hyperbolic orbit (see, for instance \cite{kat} or \cite{koz}).
The other(s) can be either hyperbolic, elliptic or parabolic.
Let us suppose that one of them is elliptic and let $s=0$ and $s=s_1$ be the arclength parameters of the trajectory.
As the motion occurs in the normal direction, 
the tangential component of the momentum $p$ is zero in both of the reflection points. 
Then $\{ (0,0),(s_1,0)\}$ is an elliptic 2-periodic orbit of the associated billiard map $T(s,p)=(S(s,p),P(s,p))$ and $(0,0)$ is an elliptic fixed point of $T^2$.

Let $R_0 = R(0)$ and $R_1 = R(s_1)$ be the radii of curvature of $\alpha$ at the 
$s=0$ and $s=s_1$ and $L=||\alpha(0)-\alpha(s_1)||$ be the length of the trajectory.
Using equation (\ref{eq:deriv}), the linear map 
$DT^2_{(0,0)} = DT_{(s_1,0)}\,\, DT_{(0,0)}$ is given by
\begin{equation}
DT^2_{(0,0)}=
\left( 
\begin{array}{cc}
\displaystyle
{\frac {(L - {R_{1}})\,(L - {R_{0}})}{{R_{1}}\,{R_{0}}} 
+ \frac {L\,(L - {R_{0}} - {R_{1}})}{{R_{0}}\,{R_{1}}}
}
& \displaystyle
{-\frac {  2\, L\,(L - {R_{1}})}{{R_{1}}}
} 
\\ 
&
\\ \displaystyle
{- \frac { 2\,(L - {R_{0}})\,(L - {R_{0}} - {R_{1}})}{{R_{0}}^{2}\,{R_{1}}}}
& \displaystyle
{\frac {(L - {R_{1}})\,(L - {R_{0}})}{{R_{1}}\,{R_{0}}}  
+ 
\frac {L\,(L - {R_{0}} - {R_{1}})}{{R_{0}}\,{R_{1}}}} 
\end{array}\right)
\label{eq:dt2}
\end{equation} 
and its eigenvalues are
$$
2\,\frac {(L - R_1)\,(L - R_0) }{R_0\,R_1} - 1
\pm
\frac{2\sqrt{L\,(L - R_0 - R_1)\,(L - R_1)\,(L - R_0)}}{R_0\,R_1}
$$
As the trajectory is elliptic the relations $L-R_0-R_1<0$ and $(L-R_0)(L-R_1)>0$ must be 
fullfilled. Assuming  that $4 \, (L-R_0)(L-R_1)\neq R_0 R_1$ and $2 \, (L-R_0)(L-R_1)\neq R_0 R_1$ then $\mu^j\neq 1$ for $j=1,2,3,4$.

In the elliptic and non-resonant case, in order to investigate the stability of the fixed point, we can proceed and examine the first Birkhoff coefficient given by (\ref{eq:tau1}) .
The complex coefficients  $c_{21}$, $c_{20}$ and $c_{02}$ in the formula depend on the real 
coefficients $a_{ij}$ and $b_{ij}$ of the Taylor expansion of $T^2$ at the origin (\ref{eq:taylor}).

The linear coefficients $a_{ij}$ and $b_{ij}$, $i+j=1$, are obviously the entries of
$DT^2_{(0,0)}$ and thus given by (\ref{eq:dt2}). 
Note that $a_{10}=b_{01}$. As $T$ is area preserving, $a_{10}^2-a_{01}b_{10}=1$ and, as 
$(0,0)$ is elliptic, $a_{01}b_{10}<0$.

These conditions were used to write down the coordinate change leading to (\ref{eq:comp}) and we  got
\footnote{http://www.mat.ufmg.br/$\sim$syok/papers/2Complex.}:
\begin{eqnarray}
&\Im (c_{21})= & {\displaystyle
\frac{a_{10}}{8}\left( -a_{21}+3\frac{b_{10}}{a_{01}}a_{03}
-3\frac{a_{01}}{b_{10}}b_{30}+b_{12}\right)
-\frac{b_{10}}{8}\left( a_{12}-3\frac{a_{01}}{b_{10}}a_{30}
- \frac{a_{01}}{b_{10}}b_{21}+3b_{03}\right) }\nonumber\\
&|c_{20}|^2 = & {\displaystyle
\frac{1}{16} \, \sqrt{ - \frac {a_{01}} {b_{10}} } \,
\left( \frac {b_{10}}{a_{01}}a_{02} + a_{20} + b_{11}  \right)^2
+ \frac{1}{16} \, \sqrt{ - \frac {b_{10}}{a_{01}}} \,
\left(  
\frac {a_{01} }{ b_{10} }b_{20} 
+ b_{02} + a_{11} \right)^2 } \\ 
&|c_{02}|^2 = & {\displaystyle
\frac{1}{16} \, \sqrt{ - \frac {a_{01}} {b_{10}} } \,
\left(\frac {b_{10} }{a_{01}}a_{02} + a_{20} - b_{11}   \right)^2
+ \frac{1}{16} \, \sqrt{ - \frac {b_{10}}{a_{01}}} \,
\left(  
\frac {a_{01} }{ b_{10} } b_{20}+ b_{02} - a_{11} \right)^2 } \nonumber
\label{eq:coef}
\end{eqnarray}
which shows that $\tau_1$ is linear on the real coefficients of third order and quadratic on the second order ones.

In order to calculate explicitly the first Birkhoff coefficient, all is needed now are the second and third order coefficients of the Taylor expansion at $(0,0)$ of
$T^2(s,p)=\left( S(S(s,p),P(s,p)),P(S(s,p),P(s,p))\right ) .$

A sequence of straightforward but long computations using the 
Chain Rule gives those Taylor coefficients. 
To illustrate it, let us give the expression of $a_{20}$
\begin{eqnarray}
a_{20}&=&\frac{\partial ^2 }{\partial s^2}S(S(s,p),P(s,p))(0,0)\nonumber \\
&=& \frac{\partial S}{\partial s}(0,0)\frac{\partial P}{\partial s}(0,0)
\frac{\partial^2 S}{\partial s\partial p}(s_1,0)
+\frac{1}{2}\frac{\partial S}{\partial s}(s_1,0)
\frac{\partial^2 S}{\partial s^2}(0,0)+\frac{1}{2}
\left[\frac{\partial P}{\partial s}(0,0)\right]^2
\frac{\partial^2 S}{\partial p^2}(s_1,0)+\nonumber \\
& &+\frac{1}{2} \frac{\partial S}{\partial p}(s_1,0)\frac{\partial^2 P}{\partial s^2}(0,0)
+\frac{1}{2}\left[\frac{\partial S}{\partial s}(0,0)\right]^2
\frac{\partial^2 S}{\partial s^2}(s_1,0)
\nonumber
\end{eqnarray}

The first derivatives of the functions $S$ and $P$ are given by formulae (\ref{eq:deriv}and they depend on the function $l(s,p)$. So, to calculate he second and third derivatives of 
$S$ and $P$ it is  necessary to evaluate the first and second derivatives of $l$.
Let $l(s,S) = || \alpha(S)-\alpha(s)||$. Then $l(s,p)= l(s,S(s,p))$.

By differentiating
$$l^2(s,S)=\left<\alpha(S)-\alpha(s),\alpha(S)-\alpha(s)\right>$$
we have
\begin{equation}
l(s,S)\,\frac{\partial l}{\partial s}(s,S)=
-\left<\alpha'(s),\alpha(S)-\alpha(s)\right> 
\label{eq:dl}
\end{equation}
and so, as $\alpha'$ is the unitary tangent vector,
$$
\frac{\partial l}{\partial s}(s,S) = - p \ .
$$
Analogously 
$$
\frac{\partial l}{\partial S}(s,S) =  P \ .
$$
These relations simply shows that $-l(s,S)$ is the generating function of the billiard map, as a Twist map.

Differentiating  (\ref{eq:dl}) with respect to $s$ and $S$
and using that $  \eta = R \, \alpha'' $ is the unitary normal vector gives
\begin{eqnarray*}
& \displaystyle {\frac{\partial^2 l}{\partial s^2}}(s,S) =& 
\frac{1-p^2}{l(s,S)} - \frac{\sqrt{1-p^2}}{R(s)} \cr
& \displaystyle{\frac{\partial^2 l}{\partial s \, \partial S}} (s,S)  = &
\frac{\sqrt{(1-p^2)(1-P^2)}}{l(s,S)} \ .
\end{eqnarray*}
The same reasoning gives
$$
\frac{\partial^2 l}{\partial S^2} (s,S)=
\frac{1-P^2}{l(s,S)} - \frac{\sqrt{1-P^2}}{R(S)}
$$
The chain rule will give, now, the first and second order derivatives of $l(s,p)$. To evaluate them at $(0,0)$ and $(s_1,0)$ it is usefull to remember that $\alpha'(s_1) = -\alpha'(0)$, $\eta(s_1) = -\eta(0)$ and $\alpha(s_1) - \alpha(0) = L \, \eta(0) $.

Because of the recurrent structure of the formulae, the explicit calculus of the $a_{ij}$ and $b_{ij}$ is suitable to be implemented as a computer program \footnote{http://www.mat.ufmg.br/$\sim$syok/papers/0ThreeJet and 1TaylorCoeffs}. The final expression of the Taylor expansion of $T^2$ is also given in the worksheet {\tt 1TaylorCoeffs}.

The second order coefficients $a_{ij},b_{ij}, i+j=2$ 
will have linear dependence on $\frac{dR}{ds}(0) = R'_0$ and $\frac{dR}{ds}(s_1) = R'_1$ while the third 
order coefficients $a_{ij},b_{ij}, i+j=3$ will have linear dependence on 
$\frac{d^2R}{ds^2}(0) = R''_0$ and $\frac{d^2R}{ds^2}(s_1) = R''_1$ and quadratic on 
the first order derivatives. 
So the first Birkhoff coefficient $\tau_1$ will be quadratic on the first derivatives of $R$ and linear on the second ones. The final expression of $\tau_1$ is obtained after substitution of the $a_{ij}$ and $b_{ij}$ into (\ref{eq:coef}) and then into (\ref{eq:tau1}) given
\begin{eqnarray*}
\tau_1 &=& -\frac{1}{8} \, 
\frac{R_0+ R_1}{R_0 \, R_1}
-\frac{1}{8} \, 
\frac{L}{L-R_0-R_1} \, 
\left(
\frac{L-R_1}{L-R_0} \, R''_0 + \frac{L-R_0}{L-R_1} \, R''_1
\right ) \\
&&
-\frac{1}{8} \, \frac{L}{(L-R_0-R_1)^2}
\left(
2 \,\frac{L-R_1}{L-R_0} \, (R'_0)^2 + 2 \,\frac{L-R_0}{L-R_1} \, (R'_1)^2
+ 3 \,R'_0 \, R'_1 
\right)\\
&&
+\frac{1}{8} \, \frac{L \, R_0 R_1}{(L-R_0-R_1)^2 \, (4\,(L-R_0)(L-R_1)-R_0 R_1)} 
\, \left(
  \frac {(L - R_1)^2 }{L - R_0}\,\frac{(R'_0)^2}{R_0}
+ \frac {(L - R_0)^2 }{L - R_1}\,\frac{(R'_1)^2}{R_1}
-  \,R'_0 \, R'_1
\right)
\end{eqnarray*}
The details leading to the formula are given in the worksheet {\tt 4Tau}.

\section{Billiards with islands}

As remarked in section \ref{sec:3}, a billiard on a strictly convex $C^2$ curve always has 2-periodic orbits and the largest one, if isolated, 
is hyperbolic. Unfortunately, one cannot assure that at least one of the others is elliptic. In fact, there are many examples where all 2-periodic orbits are
isolated and hyperbolic (see, for instance \cite{dia} or \cite{koz}). 

On the other hand, ellipticity is an open property, in the sense that if a billiard
associated to a $C^2$ strictly convex curve $\alpha$ has an elliptic 2-periodic orbit, then any strictly convex curve sufficiently $C^2$-close to $\alpha$ generates a billiard with an 
elliptic 2-periodic orbit \cite{dia}. So, a large class of strictly convex billiards has elliptic 2-periodic orbits. The question is: are they stable?

In what follows we present two classes of billiards (locally circular and symmetric)  exhibiting stable 2-periodic orbits.

\subsection{Locally Circular Billiards}

 Our first and simplest example is a 2-periodic orbit between two circles. More precisely, let $\alpha$ be a $C^5$ plane strictly convex closed curve  parameterized  by the arclength $s$, with the following properties:
\begin{list}{--}{\parsep 0pt \itemsep 5pt \topsep 0pt \partopsep 0pt}
\item  there are two points located by $s=0$ and $s=s_1$ such that 
$\alpha'(0)=-\alpha'(s_1)$ and  $\alpha(0)-\alpha(s_1)=-L\vec{\eta}(0)$, where $\vec{\eta}(0)$ 
is the unitary normal vector at $0$.
\item $\alpha$ is locally a circle, both near $s=0$ and $s=s_1$, with radii
$R_0$ and $R_1$ respectively.
\item $L$, $R_0$ and $R_1$ verify $L-R_0-R_1<0$, $(L-R_0)(L-R_1)>0$ and
$4(L-R_0)(L-R_1)\neq R_0R_1$, $2 R_0R_1$, which are open conditions on the $(L,R_0,R_1)$-space.
\end{list}
\begin{center}
\includegraphics[height=4cm,width=4.cm]{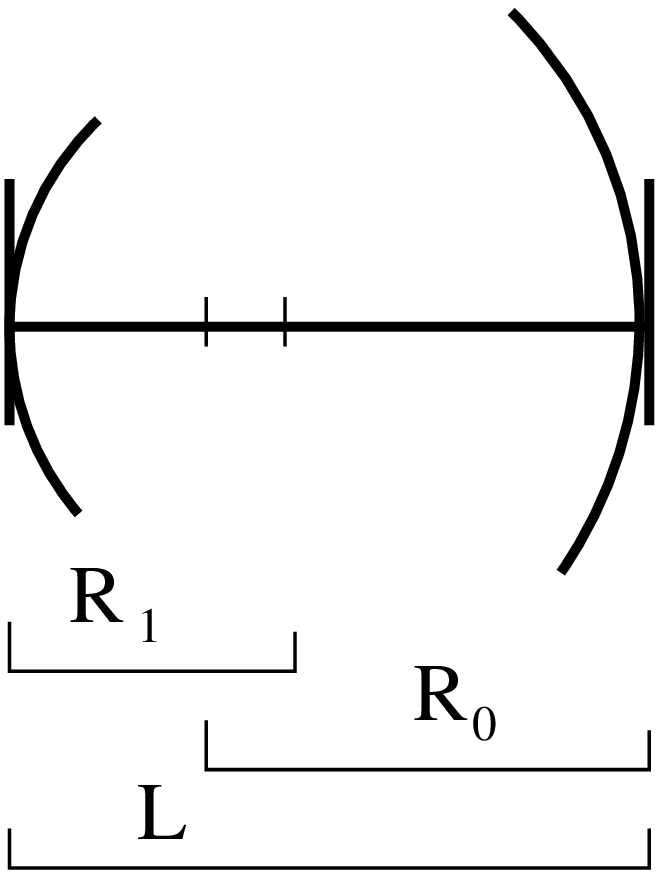}
\end{center}
With these properties, $\{(0,0),(s_1,0)\}$ is a non-resonant elliptic 2-periodic orbit
for the billiard map $T$ associated to $\alpha$. 

As $\alpha$ is locally circles, $T$ is locally analytic and the
first Birkhoff coefficient of the elliptic orbit can be calculated. 
Moreover, $R'$ and $R''$ vanish at $s=0$ and $s=s_1$.
So, 
$$\tau_1=-\frac{1}{8}\left(\frac{1}{R_0} + \frac{1}{R_1} \right)\neq 0$$
and this billiard has a stable 2-periodic orbit.

Although extremely simple, this example shows that exchanging the curve $\alpha$ by the osculating circles at the impact points gives information about ellipticity, but not about
stability, since $\tau_1$ depends on the derivatives of the radius of curvature.

\subsection{Ovals with a special symmetry}

Let $R$ be a periodic $C^4$ function with Fourier expansion
$$R(\varphi)=a_0+\sum_{n=1}a_n\cos 2n\varphi$$
with $a_n>0$ and $a_0>\sum_{n=1}a_n$, implying that 
$R(\varphi) > 0 , \forall \varphi$.

Let $\alpha$ be a curve, having $R$ as its radius of curvature, given by
$$\alpha(\varphi)=\left(x(\varphi),y(\varphi)\right)=
\left(\int_0^\varphi R(\beta)\cos\beta d\beta,
\int_0^\varphi R(\beta)\sin\beta d\beta\right).$$
It is a regular, closed and strictly convex $C^5$ curve.

\begin{center}
\includegraphics[height=4cm,width=5cm]{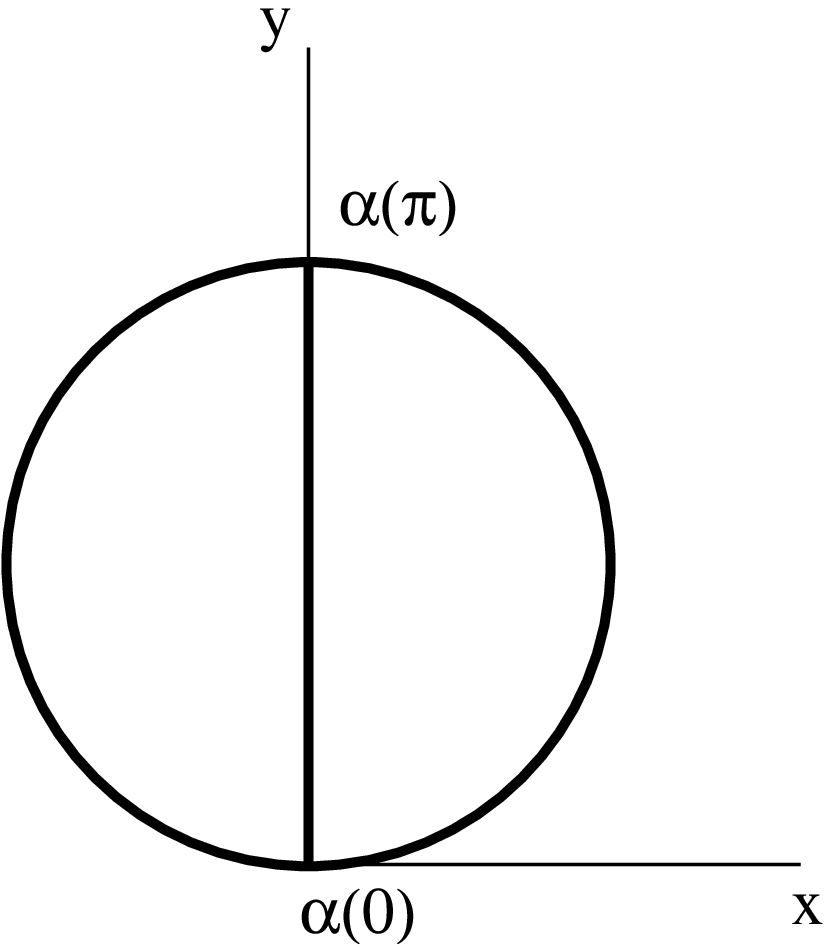}
\end{center}

As $R$ is an even function, $x(-\varphi)=-x(\varphi)$, $y(-\varphi)=y(\varphi)$ and 
$\overline{\alpha(0)\alpha(\pi)}$ is an axis of symmetry for $\alpha$. $\{(0,0),(\pi,0)\}$ is a 2-periodic orbit for the associated billiard map.

We have  
\begin{eqnarray*}
L= ||\alpha(\pi) - \alpha(0)|| = y(\pi) 
&=& 2 \, a_0 - \displaystyle{\sum_{n=1} \frac{2 \, a_n}{(2n+1)(2n-1)}} \\
R(0) = R(\pi) = R_0 &=& a_0 + \displaystyle{\sum_{n=1} a_n}
\end{eqnarray*}
and then
\begin{eqnarray*} 
L-R(0) - R(\pi) = L-2 R_0 &=&
-2 \, \sum_{n=1} a_n \left (  \frac{1}{(2n+1)(2n-1)}     + 1 \right)
< 0  \\
L-R(0) = L-R(\pi) = L-R_0 &=& 
a_0 - \sum_{n=1} a_n \left (  \frac{2}{(2n+1)(2n-1)}     + 1 \right)
\end{eqnarray*}

If 
$$
 (2+k) \, a_0 - \displaystyle{\sum_{n=1} a_n\left[\frac{4}{(2n+1)(2n-1)}+(2-k)\right]}\neq 0, \,\,\,\, k=0,\pm 1, \pm \sqrt2
$$
which are open conditions, then $\{(0,0),(\pi,0)\}$ is elliptic and non-resonant.

Let $s=s(\varphi)$ be the arclength parameter for $\alpha$. Choosing  $s(0) = 0$ and $s(\pi) = s_1$ we have  
\begin{eqnarray*} 
\left . \frac{dR}{ds} \right |_{s=0} = 
\left . \frac{1}{R_0}\frac{dR}{d\varphi} \right |_{\varphi=0} = 0
&\hbox{ , } &
\left .\frac{dR}{ds} \right |_{s=s_1}= 
\left .\frac{1}{R_0}\frac{dR}{d\varphi} \right |_{\varphi=\pi} = 0
 \\
\left .\frac{d^2R}{ds^2} \right |_{s=0} 
=
\left .\frac{1}{R_0^2}  \frac{d^2R}{d\varphi^2} \right |_{\varphi=0}
= -4\sum_{n=1}n^2 a_n< 0 &\hbox{ , }&
\left .\frac{d^2R}{ds^2}(s_1) \right |_{s=s_1}
=
\left .\frac{1}{R_0^2}\frac{d^2R}{d\varphi^2} \right |_{\varphi=\pi} = 
-4\sum_{n=1}n^2 a_n< 0
\end{eqnarray*}
and the first Birkhoff coefficient is
$$\tau_1=-\frac{1}{4R_0}\left(1+
\frac{L}{R_0(L-2R_0)}\frac{d^2R}{d\varphi^2}(0)\right)< 0.
$$
So the 2-periodic orbit is stable.

In particular, this class of curves include those studied numerically by Berry in \cite{ber} and defined by $R(\varphi) = 1+ \epsilon \cos 2 \varphi$ with $0 < \epsilon < 1$.
If $\epsilon\neq\frac{3}{5},\frac{3}{13}$ or $\frac{3}{41}(13-8\sqrt 2)$, the conditions for non-resonant ellipticity are fullfilled and the 2-periodic orbit is stable.
It is claimed in \cite{ber} that when $\epsilon=\frac{3}{5}$ (meaning parabolicity of the 2-periodic orbit) there is neutral stability. There is no specific observations for the other two values of $\epsilon$. 

It would be interesting to investigate the behaviour of this and other examples at resonances.

\bigskip

{\bf Acknowledgments} The authors would like to thank M. J. Dias Carneiro for many enlightening discussions. This work is supported by CNPq and FAPEMIG brazilian agencies.

\end{document}